%% file: main.tex
\documentclass[11pt]{article}
\usepackage[T1]{fontenc}
\usepackage[utf8]{inputenc}
\usepackage[english]{babel}
\usepackage[margin=1in]{geometry}
\usepackage{amsmath,amssymb,amsthm}
\usepackage{newtxtext,newtxmath}
\usepackage{microtype}
\usepackage{graphicx,booktabs,algorithm,algpseudocode}
\usepackage[round,authoryear]{natbib}
\usepackage{xurl}
\usepackage[hidelinks]{hyperref}
\newtheorem{proposition}{Proposition}

\DeclareMathOperator{\Cov}{Cov}
\DeclareMathOperator{\Var}{Var}
\newcommand{\E}{\mathbb{E}}
\newcommand{\CS}{\mathcal{C}_{\alpha}}
\newcommand{\Reg}{\mathcal{R}}

\title{Approximate minimax regret for conditionally similar tests\thanks{The author thanks Isaiah Andrews for helpful comments.}}
\author{Andr\'es Aradillas Fern\'andez\textsuperscript{a}\thanks{Corresponding author. Email: \href{mailto:aradilla@mit.edu}{aradilla@mit.edu}.}\\[4pt] \small\textsuperscript{a}Department of Economics, Massachusetts Institute of Technology\\ \small 77 Massachusetts Avenue, Building E52-502\\ \small Cambridge, MA 02139, United States}
\date{\today}

\begin{document}
\maketitle
\begin{abstract}
We construct approximately minimax regret optimal tests in a scalar Gaussian instrumental variables experiment. This example shows how conditionally similar tests allow a novel approximation approach, allowing us to optimize weights over alternatives to minimize worst-case power regret. Stochastic mirror descent and averaging yield finite-iteration guarantees with exact conditional similarity. We establish bounds for exact and simulated powers and illustrate the procedure numerically. This initial algorithm shows how existing nearly-optimal test computation methods can be extended to include composite alternative hypothesis sets without the computational cost associated with maintaining size control.
\end{abstract}

\section{Introduction}

Conditional similarity controls rejection probabilities in the presence of nuisance parameters but generally leaves many tests to choose from \citep{moreira2003conditional}. We select among these tests by minimizing their largest power loss relative to the conditionally similar power envelope. This criterion is useful when instruments may be weak and no single test maximizes power at every alternative.

Our contribution is to use conditional size control to optimize power regret across alternatives. \citet{elliott2015nearly}, henceforth EMW, fix a weighting distribution over alternatives; weighted average power then equals power against a single mixture distribution. Their numerical search concerns a least-favorable distribution under the null, and thus the algorithm seeks to find an optimal combination of tests that satisfy these size constraints simultaneously. While Section 4.3 of EMW does account for composite alternative hypothesis sets, our conditioning construction allows for a save on compute time, as we maintain size guarantees by construction. \citet{aradillas2025approximate} then develop stochastic mirror descent (SMD) for the corresponding finite-null, single-alternative problem, thus attaining approximation guarantees not present in the algorithm of EMW. That formulation also accommodates a fixed alternative mixture representing a weighted-average-power (WAP) criterion (for a discussion, see Section 3.8 of \cite{lehmann2005testing}), restricting the problem to that of a composite null and a singleton alternative.

We instead impose conditional similarity within every best response. This removes the need to optimize null weights to enforce size and allows the iterative search to focus on alternatives at which power falls furthest below its conditionally similar envelope. How \cite{aradillas2025approximate} construct their algorithm is by utilizing the subgradient constructed from the Lagrangian, but by freeing up this size constraint, we are instead able to look at many combinations of alternate hypotheses. Weighted average power remains the best-response objective, but its weights are updated to solve a minimax regret problem rather than specified in advance as the final power criterion. Our result concerns the conditionally similar class in the Gaussian experiment, instead of the unrestricted class of level-$\alpha$ tests, but this flexible method may lend itself to tractable results beyond Gaussian settings, which is accomplished in \cite{aradillas2025approximate} for the single-alternative version of the problem.

For a finite alternative grid, we combine conditional Neyman--Pearson best responses with exponential weight updates, the latter of which help preserve the geometry of the probability simplex. Averaging preserves exact conditional similarity and yields explicit regret bounds for exact and simulated power calculations.

The contribution of this work lies in the conditionally similar testing literature. \citet{andrews2016conditional} studies minimax regret within a class of conditional linear combination tests, while \cite{chernozhukov2009admissible} and later \citet{olea2020admissible} characterize admissible similar tests. Not yet established in the literature is how to compute minimax regret optimal tests, which is the main contribution of this work, as we optimize over the full conditionally similar class here. Furthermore, we connect to existing literature such as \cite{andrews2006optimal} that seeks to understand the power envelope for an invariant similar class exactly in the Gaussian example that we consider.

\section{Setup and Algorithm}
\subsection{Testing problem}

As in a weak instruments setting, we observe a pair of reduced-form estimates satisfying
\begin{equation}\label{eq:experiment} \begin{pmatrix}\hat\delta\\\hat\pi\end{pmatrix}\sim N\!\left(\begin{pmatrix}\pi\beta\\\pi\end{pmatrix},\Sigma\right),\qquad \Sigma \text{ positive definite}, \end{equation}
where $\Sigma$ is known and does not depend on $(\beta,\pi)$. We test $H_0:\beta=\beta_0$ at level $\alpha\in(0,1)$, with unrestricted nuisance parameter $\pi\in\mathbb R$. Define
\begin{equation}\label{eq:transformation} S=\hat\delta-\beta_0\hat\pi,\qquad a=\Cov(\hat\pi,S)/\Var(S),\qquad U=\hat\pi-aS. \end{equation}
The transformation is invertible. Let $\sigma_S^2=\Var(S)$ and $\sigma_U^2=\Var(U)$. At $\theta=(\beta,\pi)$, the variables $S$ and $U$ are independent normals with means $m(\theta)=\pi(\beta-\beta_0)$ and $\tau(\theta)=\pi-am(\theta)$ and variances $\sigma_S^2$ and $\sigma_U^2$. Under the null, $S\mid U=u\sim N(0,\sigma_S^2)$, independently of $\pi$.

A randomized test is a measurable rejection probability $\varphi(s,u)\in[0,1]$. Writing $f_0$ for the $N(0,\sigma_S^2)$ density, we consider
\begin{equation}\label{eq:class} \CS=\left\{\varphi:\int\varphi(s,u)f_0(s)\,ds=\alpha\ \text{for Lebesgue-a.e. }u\right\}. \end{equation}
Because the null density of $U$ is everywhere positive, this restriction is equivalent to conditional similarity for every $\pi$. It implies unconditional rejection probability $\alpha$.

For alternatives $\theta\in\Theta\subseteq\{(\beta,\pi):\beta\ne\beta_0\}$, define the power envelope, regret, and minimax value by
\begin{equation}\label{eq:regret} b(\theta)=\sup_{\varphi\in\CS}\E_\theta[\varphi],\qquad \Reg(\varphi,\theta)=b(\theta)-\E_\theta[\varphi],\qquad V_\Theta=\inf_{\varphi\in\CS}\sup_{\theta\in\Theta}\Reg(\varphi,\theta). \end{equation}
We first take $\Theta=\Theta_I=\{\theta_1,\ldots,\theta_I\}$, with $I\ge2$, and write $V_I=V_{\Theta_I}$ and $b_i=b(\theta_i)$.

\subsection{Conditional best responses and weight updates}

For weights $w$ in the probability simplex $\Delta^{I-1}$, let $g_w=\sum_iw_i g_i$, where $g_i$ is the joint density of $(S,U)$ at $\theta_i$. The best response maximizes weighted average power:
\begin{equation}\label{eq:bestresponse} \varphi_w\in\arg\max_{\varphi\in\CS}\sum_{i=1}^I w_i\E_{\theta_i}[\varphi]. \end{equation}
The weights $w$ are optimization variables for worst-case regret. Thus each best response uses a single alternative mixture, but the mixture changes across iterations as the algorithm reallocates weight toward high-regret alternatives. The constraint in \eqref{eq:class} separates across $u$. Conditional Neyman--Pearson optimization therefore rejects for large values of
\begin{equation}\label{eq:ratio} L_w(s,u)=\frac{g_w(s,u)}{f_0(s)}=\sum_{i=1}^I w_i f_U(u;\tau_i)\exp\!\left(\frac{m_i s}{\sigma_S^2}-\frac{m_i^2}{2\sigma_S^2}\right), \end{equation}
where $m_i=m(\theta_i)$, $\tau_i=\tau(\theta_i)$, and $f_U(\cdot;\tau_i)$ is the $N(\tau_i,\sigma_U^2)$ density. Specifically, $\varphi_w=\mathbf1\{L_w>c_w(u)\}+\gamma_w(u)\mathbf1\{L_w=c_w(u)\}$, with threshold and boundary randomization chosen to satisfy \eqref{eq:class}. For a single alternative, the optimal rule is one-sided in $S$, giving
\begin{equation}\label{eq:envelope} b_i=\Phi\!\left(\Phi^{-1}(\alpha)+|m_i|/\sigma_S\right), \end{equation}
where $\Phi$ is the standard normal CDF. When $m_i=0$, the envelope equals $\alpha$.

The function $F(w)=\sup_{\varphi\in\CS}\sum_iw_i(\E_{\theta_i}[\varphi]-b_i)$ is convex, with subgradient $G_i(w)=\E_{\theta_i}[\varphi_w]-b_i$. Negative-entropy mirror descent thus increases weights in proportion to estimated regret (see Section 4.2 of \cite{bubeck2015convex}, as well as \cite{freund1999adaptive}, after which this algorithm is modeled). Starting from $w_{1,i}=1/I$, at iteration $k$ we compute $\varphi_k=\varphi_{w_k}$ and set
\begin{equation}\label{eq:update} \widehat p_{k,i}=\frac1N\sum_{n=1}^N\varphi_k(Y_{k,i,n}),\qquad \widehat G_{k,i}=\widehat p_{k,i}-b_i,\qquad w_{k+1,i}=\frac{w_{k,i}\exp(-\eta\widehat G_{k,i})}{\sum_jw_{k,j}\exp(-\eta\widehat G_{k,j})}. \end{equation}
Conditional on the preceding iterations, the draws $Y_{k,i,n}\sim g_i$ are independent across $i$ and $n$ and fresh at each iteration. Hence $\widehat G_k$ is conditionally unbiased. Exact-power updates replace $\widehat p_{k,i}$ by $\E_{\theta_i}\varphi_k$.

We return $\bar\varphi_K=K^{-1}\sum_{k=1}^K\varphi_k$, which could also be implemented by selecting an iteration uniformly and applying its test. Note that, conditional on simulation history, $\bar\varphi_K\in\CS$ by construction. Thus simulation affects regret but does not affect conditional size when best responses are computed exactly. Algorithm \ref{alg:smd} specifies the full implementation.

\begin{algorithm}[t]
\caption{Conditionally similar testing by SMD}\label{alg:smd}
\begin{algorithmic}[1]
\Require Grid $\Theta_I$, size $\alpha$, tolerance $\epsilon$, failure probability $\delta$, draws $N$.
\State Choose $K$ and $\eta$ according to \eqref{eq:iterations}.
\State Compute envelopes $b_i$ using \eqref{eq:envelope}; set $w_{1,i}=1/I$.
\For{$k=1,\ldots,K$}
\State Construct $\varphi_k$ from \eqref{eq:ratio}, imposing conditional size $\alpha$.
\State Draw fresh $Y_{k,i,n}\sim g_i$ independently for all $i,n$.
\State Set $\widehat p_{k,i}=N^{-1}\sum_n\varphi_k(Y_{k,i,n})$ for every $i$.
\State Set $w_{k+1,i}\propto w_{k,i}\exp\{\eta(b_i-\widehat p_{k,i})\}$; normalize.
\EndFor
\State \Return $\bar\varphi_K=K^{-1}\sum_k\varphi_k$.
\end{algorithmic}
\end{algorithm}

\subsection{Algorithm guarantees}

\begin{proposition}\label{prop:bounds}
For the exact conditional best responses in \eqref{eq:bestresponse}, the following statements hold for every $K\ge1$ and $\eta>0$.
\begin{enumerate}
\item With exact-power updates,
\begin{equation}\label{eq:deterministic} \max_i\Reg(\bar\varphi_K,\theta_i)\le V_I+\frac{\log I}{\eta K}+\frac{\eta}{8}. \end{equation}
\item With the simulated updates in \eqref{eq:update}, for any $\delta\in(0,1)$, with probability at least $1-\delta$ over the simulation draws,
\begin{equation}\label{eq:stochastic} \max_i\Reg(\bar\varphi_K,\theta_i)\le V_I+\frac{\log I}{\eta K}+\frac{\eta}{2}+\sqrt{\frac{2\log((I+1)/\delta)}{NK}}. \end{equation}
\end{enumerate}
\end{proposition}

With exact powers, $\eta=\sqrt{8\log I/K}$ gives error $\sqrt{\log I/(2K)}$. With simulated powers, an error tolerance $\epsilon>0$ and failure probability $\delta$ are attained by
\begin{equation}\label{eq:iterations} K=\left\lceil\epsilon^{-2}\left(\sqrt{2\log I}+\sqrt{\frac{2\log((I+1)/\delta)}N}\right)^2\right\rceil,\qquad \eta=\sqrt{\frac{2\log I}{K}}. \end{equation}
The guarantees concern the average test, not the best response to average weights, which may underperform (e.g., see the discussion surrounding Algorithm 1 of \cite{aradillas2025approximate}). Note that these results do not require a minimax interchange; weighted regret is bounded by $V_I$ directly. Indeed, \eqref{eq:ew} also establishes $V_I=-\min_{w\in\Delta^{I-1}}F(w)$ by letting $K$ increase, combining weak duality with the bound on average weighted regret. The continuity of $F$ ensures attainment.





\section{A numerical illustration}\label{sec:example}

We set $\beta_0=0$, $\Sigma=I_2$, and $\alpha=0.05$, so that $S=\hat\delta$ and $U=\hat\pi$. Our grid contains $\theta_1=(-3,0.5)$ and $\theta_2=(1,2)$, giving $(m_1,m_2)=(-1.5,2)$ and $(\tau_1,\tau_2)=(0.5,2)$. The corresponding power envelopes are $0.4424$ and $0.6388$. This deliberately small grid isolates the trade-off between power against negative and positive effects.

For these two alternatives, the likelihood ratio in \eqref{eq:ratio} is strictly convex in $s$. The best response rejects in two tails. If $q_w(u)$ is its left-tail null probability, the critical values satisfy
\begin{equation}\label{eq:tailallocation} \ell_w(u)=\Phi^{-1}(q_w(u)),\qquad h_w(u)=\Phi^{-1}(1-\alpha+q_w(u)),\qquad L_w(\ell_w(u),u)=L_w(h_w(u),u). \end{equation}
We solve the last equality by bisection over $q\in[10^{-14},\alpha-10^{-14}]$. The resulting rule has conditional size $q+(\alpha-q)=\alpha$ by construction. Larger values of $u$ favor the second alternative and shift rejection probability toward the right tail.

We set the target excess regret to $\epsilon=0.02$, the failure probability to $\delta=0.05$, and the number of draws per alternative per iteration to $N=512$. Equation~\eqref{eq:iterations}, with $I=2$, then selects $K=4{,}251$ and $\eta=0.01806$. Under exact conditional best responses, Proposition~\ref{prop:bounds} gives probability at least $0.95$ that the average test's maximum grid regret is at most $V_I+0.02$. We evaluate power by integrating the conditional rejection probability with 320-node Gauss--Hermite quadrature. Doubling the quadrature order from 160 to 320 changes computed iteration-specific powers by less than $10^{-12}$, suggesting numerical error is very low here. The reported values then come from one simulation run.

\begin{table}[t]
\centering
\caption{Power and regret on the two-alternative grid}\label{tab:example}
\small
\input{results/example_table.tex}
\par\smallskip
\begin{minipage}{\linewidth}
\footnotesize\textit{Notes:} Size for all cases is $\alpha = .05$. The equal-tail test rejects when $|S|>\Phi^{-1}(0.975)$. The initial best response uses $w=(1/2,1/2)$. SMD rows average all tests through the indicated iteration of the same run. Powers are evaluated by numerical integration; size follows from the tail allocation. Maximum regret is relative to the pointwise envelopes in \eqref{eq:envelope}.
\end{minipage}
\end{table}

\begin{figure}[t]
\centering
\includegraphics[width=0.72\linewidth]{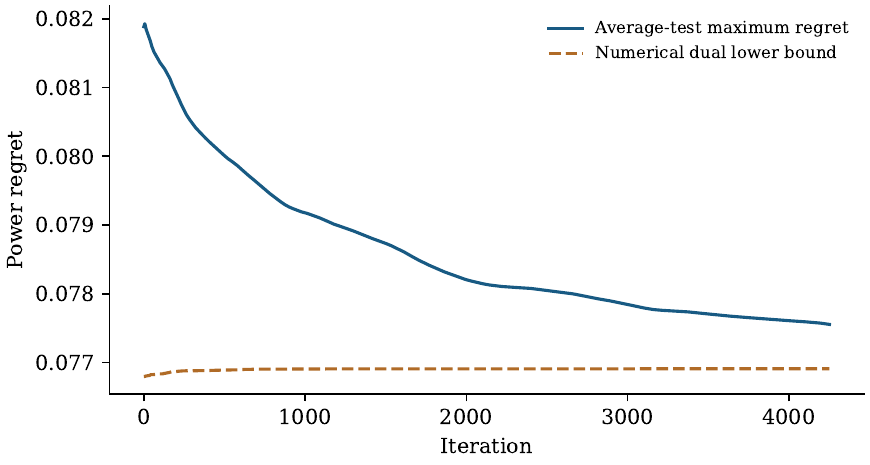}
\caption{Regret along the SMD iterations. We report the maximum regret of the average test and the running maximum of $w_k^\top r_k$, evaluated by numerical integration. With exact best responses and integration, the latter bounds $V_I$ from below.}\label{fig:example}
\end{figure}

\begin{figure}[t]
\centering
\includegraphics[width=0.72\linewidth]{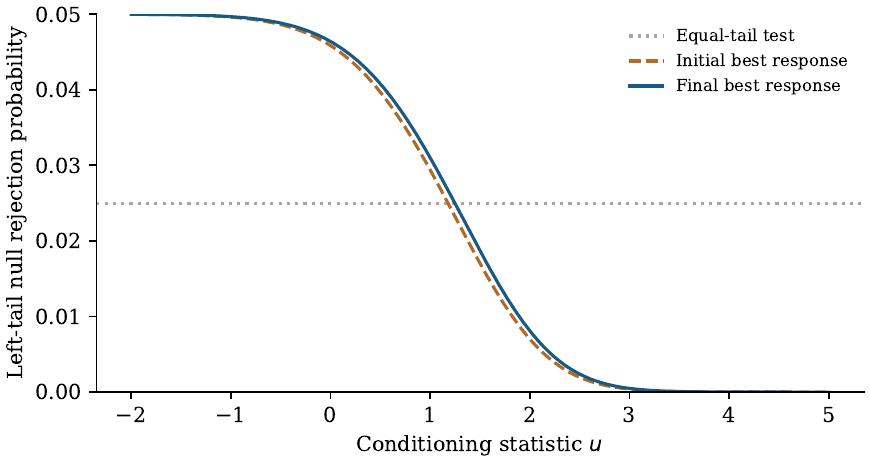}
\caption{Conditional tail allocation. Left-tail null rejection probabilities for the initial and final individual best responses are compared with the equal-tail test.}\label{fig:tails}
\end{figure}

Table~\ref{tab:example} shows that conditioning improves power at both alternatives relative to the equal-tail test. Figure~\ref{fig:tails} shows how the conditioning statistic shifts rejection probability between tails. Subsequent weight updates increase power at $\theta_1$ while sacrificing some power at $\theta_2$, reducing the largest regret. After $4{,}251$ iterations, maximum regret is $0.0776$, compared with $0.0819$ initially and $0.1228$ for the equal-tail test. Figure~\ref{fig:example} reports a numerical dual lower bound of $0.0769$, leaving a numerical gap of $0.00065$. The small distance between the initial best response and the output of Algorithm \ref{alg:smd} suggests the example here could benefit from early stopping, but recall that Proposition \ref{prop:bounds} is only a worst-case guarantee. Furthermore, monotonicity in average-test maximum regret is not always guaranteed, and in fact between the first and second iterations, regret goes up in our example. The dual lower bound therefore only acts as a guide on the maximum amount by which minimax regret can be reduced.

\section{Conclusion}

Conditional similarity allows us to shift the numerical search from null distributions enforcing size to alternative distributions determining worst-case power regret. Relative to EMW's fixed WAP criterion, the alternative weights are chosen endogenously to balance regret within the conditionally similar class, allowing robustness to WAP misspecification. Conditional Neyman--Pearson best responses and SMD deliver approximation guarantees while preserving exact conditional size. This work can serve the basis for future work moving beyond the Gaussian setup of the problem, with the built-in size control allowing for optimization over many alternative hypotheses.

\section*{Data and code availability}

We use simulated data. The reproducible Python package, as well as instructions, are available at \url{https://github.com/aaf101/SMD-for-Conditional-Similarity}.

\section*{Declaration of generative AI and AI-assisted technologies in the manuscript preparation process}

During the preparation of this work, the author used GPT-6 to help with the writing, proof-checking, and code implementation in this paper. The author directed this work and subsequently reviewed and verified all output. The author takes full responsibility for the content of the article.

\appendix
\section{Proofs of Main Results}

\subsection{Proof of Proposition \ref{prop:bounds}}

This proof follows a standard Hedge derivation. Write $r_{k,i}=\Reg(\varphi_k,\theta_i)\in[0,1]$. Best-response optimality gives $w_k^\top r_k=\inf_{\varphi\in\CS}\sum_iw_{k,i}\Reg(\varphi,\theta_i)\le V_I$. The exponential-weights inequality for rewards in $[0,1]$ (Theorem 2.2 of \cite{cesa2006prediction}) yields
\begin{equation}\label{eq:ew} \max_i\frac1K\sum_kr_{k,i}\le\frac1K\sum_kw_k^\top r_k+\frac{\log I}{\eta K}+\frac\eta8. \end{equation}
Linearity of power proves \eqref{eq:deterministic}.

For simulated updates, $\widehat G_{k,i}\in[-1,1]$. The same inequality, now with range length two, gives, simultaneously for all $i$,
\begin{equation}\label{eq:noise} \frac1K\sum_k(r_{k,i}-w_k^\top r_k)\le\frac{\log I}{\eta K}+\frac\eta2+\frac1K\sum_k\Delta_{k,i}-\frac1K\sum_kw_k^\top\Delta_k, \end{equation}
where $\Delta_{k,i}=\widehat G_{k,i}-G_i(w_k)$. Conditional Hoeffding bounds and iteration over the simulation history imply $\Pr(K^{-1}\sum_k\Delta_{k,i}>x)\le \exp(-2NKx^2)$. The same bound holds for $-K^{-1}\sum_kw_k^\top\Delta_k$: conditional independence gives squared summand ranges totaling $\sum_iw_{k,i}^2/N\le1/N$ at each iteration. A union bound over these $I+1$ deviations, with $x=\sqrt{\log((I+1)/\delta)/(2NK)}$, bounds the two noise terms in \eqref{eq:noise} by $2x$. Using $w_k^\top r_k\le V_I$ and linearity proves \eqref{eq:stochastic}.

\bibliographystyle{apalike}
\bibliography{references}
\end{document}

%% file: results/example_table.tex
\begin{tabular}{lrrr}
\toprule
Test & Power at $\theta_1$ & Power at $\theta_2$ & Max. regret \\
\midrule
Equal-tail test & 0.323 & 0.516 & 0.123 \\
Initial best response & 0.361 & 0.567 & 0.082 \\
SMD average, 100 iterations & 0.361 & 0.567 & 0.081 \\
SMD average, 4251 iterations & 0.365 & 0.563 & 0.078 \\
\bottomrule
\end{tabular}